\documentclass[twocolumn,prl]{revtex4}

\usepackage{graphicx}%
\usepackage{dcolumn}
\usepackage{amsmath}
\usepackage{psfrag}
\begin{document}

\preprint{cond-mat/none so far}

\title[Short Title]{Growing Networks with Enhanced Resilience to Perturbation}
\author{Markus Brede}
\email{Markus.Brede@Csiro.au}
\author{John Finnigan}%
\affiliation{%
CSIRO Centre for Complex Systems, Canberra, ACT 2601, Australia
}%

\date{5.5.2004}
\begin{abstract}
\noindent Scale-free (SF) networks and small world  networks have been found to occur in very diverse contexts. It is this striking universality which makes one look for widely applicable mechanisms which lead to the formation of such networks. In this letter we propose a new mechanism for the construction of SF networks: Evolving networks as interaction networks of systems which are distinguished by their stability if perturbed out of equilibrium. Stability is measured by the largest real part of any eigenvalue of a matrix associated with the graph. We extend the model to weighted directed networks and report power law behaviour of the link strength distribution of the weighted graphs in the SF regime. The model we propose for the first time relates SF networks to stability properties of the underlying dynamical system.

\end{abstract}
\pacs{Valid PACS appear here}

\maketitle

Recent studies have shown that a SF topology of the interaction network is a universal feature shared by many complex coupled systems. Examples are found in diverse fields including the WWW, traffic flow systems, social networks and  genetic, metabolic, and protein folding networks \cite{handbook}. Several mechanisms for the formation of SF networks are known. First SF networks can be built by preferential attachment \cite{Bar1} where new nodes form links preferentially to old nodes of high degree. Modifications to this procedure --- e.~g., incorporating an {\it a priori} assigned individual node fitness --- are still based on the general mechanism of peferential attachment and result in a slightly modified network topology \cite{Bianconi}. Further, SF networks can be considered as the result of a process optimizing the diameter and link number in a network of given size \cite{Sole}, the direct construction of Hamiltonians for the network \cite{Berg} or a thresholding mechanism \cite{Capocci}. Very recently also, another mechanism via node merging was discovered \cite{Sneppen}.

It has been shown that SF networks are very robust to random node removal \cite{Bar2}, thus allowing speculation that during evolution SF topologies might have been selected because of the inherent stability associated with their architectures. However, stability measures in the above work have been purely topological. The real situation appears far more complicated. In most systems, the network topology only reflects the population dynamics as given by an underlying set of simultaneous equations. Heterogeneity in link strengths (denoting the strength of couplings in the equations) further complicates the picture.


Here, by directly relating a system's stability to its topology, we propose another mechanism to obtain SF networks. The mechanism we use comprises two essential steps: (i) random addition of new nodes to the network and (ii) selection of more stable networks, where stability is measured by the size of the largest eigenvalue of a matrix associated with the graphs.

Consider a complex system the dynamics of which are described by some set of non-linear first order differential equations. If the system approaches an equilibrium state, a linear stability analysis can be undertaken, yielding
\begin{align}
\dot{x} = M x,
\end{align}
where $x$ denotes deviations from equilibrium and $M$ stands for the Jacobian matrix at equilibrium. The equilibrium is stable, if the largest real part $\lambda_\text{max}$ of any eigenvalue of $M$ is less than zero, where $\lambda_\text{max}=\text{max}_{\lambda \in \sigma(M)} \text{Re}(\lambda)$ and $\sigma(M)$ is the spectrum of $M$. For generality we do not specify the exact form of these equations underlying the dynamics of the network, but concentrate on the Jacobian alone, following an approach similar to \cite{May,OIKOS}.

The main diagonal elements of $M$ are set to $m_{i i}=-1$, thus the populations are self regulated and normalized with respect to their intrinsic growth rates. Non-diagonal elements of $M$ give the adjacency matrix of the network. We start allowing only matrix elements $m_{i j}\in \{-1,+1\}$, $m_{i j}=-1$ representing suppresion of node $i$'s population by $j$ and $m_{i j}=+1$ a stimulation to growth.

\begin{figure} [tbp]
\begin{center}
 \includegraphics [width=8cm]{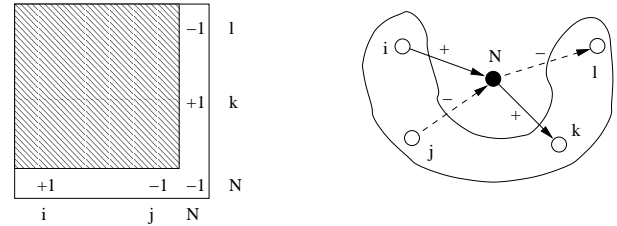}
 \caption {The attachment process. Dashed lines indicate negative links, solid lines positive links. As described in the text a negative link has strength -1, a positive link strength +1. On the right hand side it is visualized how the new node links into the old network. It always forms one positive and one negative in- and outlink with 4 randomly chosen vertices of the old network. The left hand figure shows how the attachment to the network changes the matrix $M$.}
 \label{F0}
\end{center}
\end{figure}
To proceed, we propose a mechanism for the growth of a directed network with positive and negative connections based on the $\lambda_\text{max}<0$  criterion. We start with a disconnected set of $N_0=4$ nodes, hence $m_{i j}=-\delta_{i j}$, $i,j=1,...,4$ and $\lambda_\text{max}(4)=-1$, and continue with the iteration of the following steps:
(i) Add a new node to the network, which forms $l=2$ positive connections to randomly selected nodes, of which one is an in-link and the other an out-link. Next we add another $l=2$ negative links in the same fashion [see Fig. \ref{F0}].
(ii) The eigenvalues of $M$ are determined. Let $\lambda_\text{max}(N)$ denote the largest real part of all eigenvalues of $M$, and let $\lambda_\text{max}(N-1)$ be the same for the $(N-1)\times (N-1)$ matrix before insertion of the new node. If $\lambda_\text{max} (N)\geq 0$ the configuration is rejected immediately and we proceed with (i) and the last accepted configuration.
(iii) If $\lambda_\text{max}(N)<\lambda_\text{max}(N-1)$ the last node addition will be accepted. Otherwise, the acceptance probability is given by
\begin{align}
p_\text{accept}= \exp \left(-\beta (\lambda_\text{max}(N)-\lambda_\text{max}(N-1)) \right),
\label{Eq_acc}
\end{align}
where $\beta$ is an inverse temperature-like parameter.
Unless the desired network size $N$ has been reached, the algorithm continues with step (i). If before reaching the target network size a configuration to which no further node can be added is encountered, we start again from step (i).

By construction every matrix $M$ has the same number of positive and negative links. Furthermore, $\text{Tr} (M)=\sum_i \lambda_i=-N$ giving $\overline{\lambda}=1/N \text{Tr} M=-1$. Hence, in a stable matrix always $\text{Re} (\lambda) \in (0,-N)$ and, in particular, $\lambda_\text{max} \in (0,-1]$. Consequently, any broadening of the eigenvalue distribution in the direction of smaller eigenvalues will also entail a larger eigenvalue closer to zero and thus a less stable system in our sense. Note, that by defining a Markov process in $\lambda_\text{max}(N)$ the above procedure also describes a Markov process in the space of all graphs with positive and negative links. The system size $N$ gives the number of steps the walk has to perform.


The algorithm can be interpreted in two ways. First, one may consider it only as a method to construct an ensemble of matrices (or graphs) with optimized stability properties. In this interpretation, the largest eigenvalue of a matrix might be considered as its `energy', $\beta$ being a measure for the fluctuations in the ensemble. Neglecting the system's growth, the procedure is then very similar to a Metropolis algorithm as commonly used for the numerical study of spin systems.

On the other hand, it could also be conceived as a network evolution, mimicking a system's growth one node at a time. Every given interval of time a new `species' is added to the system. This causes a perturbation to the population dynamics, that again settles into an equilibrium. After relaxation, thermal fluctuations lead to perturbations around the equilibrium which cause the less stable systems to collapse. In this view, the algorithm describes an evolutionary search, letting only the most stable systems survive.

{\it Degree distributions--} In order to compare the results obtained by the algorithm to a random network evolution, we calculate the degree distribution of the network which would be obtained by only iterating step (i) of the algorithm. For simplicity we don't distiguish between positive and negative links in the evaluation of degree distributions in this letter. Following the rate equation approach \cite{Redner} one quickly obtains for nodes of degree larger than one (nodes of degree smaller than two remain from the initial conditions and don't affect the asymptotic limit)
\begin{align}
 \langle n_i(t+1)\rangle = \langle n_i(t)\rangle -\langle 2/N(t)n_i(t)\rangle+
\langle 2/N(t)n_{i-1}(t)\rangle,
\label{Eq2}
\end{align}
where $n_i(t)$ stands for the number of nodes of degree $i$ after iteration $t$. Assuming $\langle n_i(t)\rangle \sim p_iN(t)$ in the limit of large network sizes, Eq.'s (\ref{Eq2}) yield an asymptotic degree distribution
\begin{align}
 p_i= \frac{1}{2} \left(\frac{2}{3}\right)^{i-1},
\label{rand_deg}
\end{align}
which is exponential. We now continue by comparing Eq. (\ref{rand_deg}) with degree distributions obtained in the ensemble of graphs constructed by the above algorithm.
\begin{figure} [tbp]
\begin{center}
\includegraphics [width=7cm]{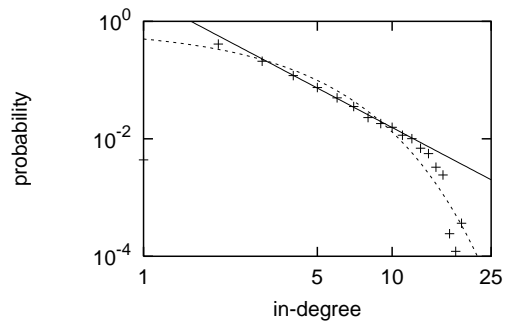}
 \caption {Example for the in-degree distribution of graphs ($N=100$) constructed with the above algorithm. The solid line (note the log scale on both axes) indicates a power law with exponent $\gamma \approx -2.23\pm 0.05$. For comparison, the dashed line shows an exponential network as given by Eq. (\ref{rand_deg}). The data are sampled for $\beta=100$, $N_0=4$, and represent averages over $1000$ independent runs.}
 \label{F1}
\end{center}
\end{figure}
Figure \ref{F1} shows simulation results for the in-degree distribution of networks of size $100$ constructed with $\beta=100$. Up to a finite size cut-off both the in- and out-degree distributions follow the same power law $\langle n_i\rangle\sim i^{-\gamma}$ with an exponent $\gamma_\text{in}=\gamma_\text{out}=\gamma=-2.23\pm0.05$.\\ {\it The influence of $\beta$--} Assuming a fixed network size the most important parameter in our construction is $\beta$. To quantify the influence of $\beta$ on the degree distribution we define a degree entropy $S_\text{deg}=\sum_i p_i \log p_i$ (cf. Ref. \cite{Sole}).
\begin{figure} [tbp]
\begin{center}
\psfrag{exponential}{exponential}
\psfrag{networks} {networks}
\psfrag{scale-free}{scale-free}
\psfrag{mixed configurations} {`mixed configurations'}
\psfrag{beta}{{\large $\beta$}}
\psfrag{S}{{\large $S_\text{deg}$}}
\psfrag{deg} {}
\includegraphics [width=7cm] {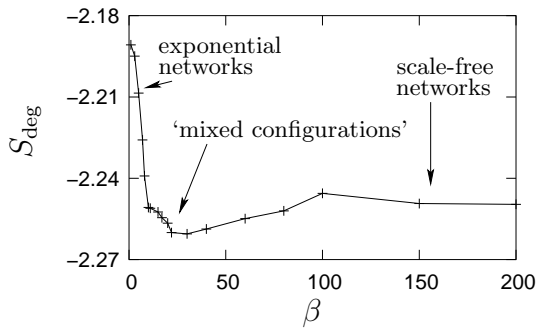}
 \caption {Simulation data showing changes in the degree entropy $S_\text{deg}$ with $\beta$ for a system of size $N=50$. We find, that the relatively high degree entropies for small $\beta$ correspond to exponential networks, while all networks constructed for high $\beta$ are SF. In the intermediate region, a subensemble of the networks is SF, while some networks are neither SF nor exponential.}
\label{F1a}
\end{center}
\end{figure}
\begin{figure} [tbp]
\psfrag{l} {$\lambda_\text{max}$}
\psfrag{max} {}
\psfrag{l1} {$\lambda_\text{c}$}
\begin{center}
\includegraphics [width=7cm] {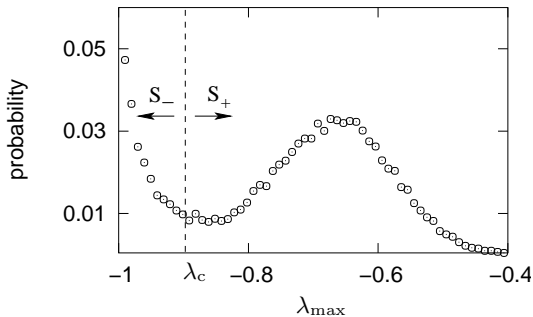}
 \caption {Histogram counting the normalized frequency of how many times networks with $\lambda_\text{max}$ occur if constructing $10^4$ networks of size $50$. The distribution is bimodal with peaks at $\lambda=-1$ and $\lambda\approx -.65$. We find that networks around the peak at $\lambda_\text{max}=-1$ are SF.}
\label{F1b}
\end{center}
\end{figure}

Figure \ref{F1a} illustrates changes in network structure with $\beta$. For very small $\beta$ we find networks with an exponential degree distribution, which is well discribed by Eq. (\ref{rand_deg}). On the opposite end, for high $\beta$, networks are SF. In the intermediate range, both SF and non-SF networks occur. A more detailed investigation reveals, that the transition between exponential and SF networks is related to a shift in typical largest real parts of eigenvalues. To illustrate this, Fig. \ref{F1b} displays data for the distribution of $\lambda_\text{max}$. This distribution is bimodal. Corresponding to the two peaks we define two ensembles of networks, $S_-=\{M\in S|\lambda_\text{max}(M)<-\lambda_\text{c}\}$, $\lambda_\text{c}=-.9$, (the more stable subset) and $S_+=S-S_-$ (the less stable subset). Interestingly, independent of $\beta$, networks belonging to the more stable subset are SF while others belonging to $S_+$ are typically not SF. A further finding is, that the degree distributions of the less stable networks are more likely to be exponential the farther away the second peak is from $\lambda=-1$.

In the following we relate changes in $\beta$ to changes in the weight of both peaks in the $\lambda_\text{max}$-distribution and thus to changes in network topology. Choosing a high value of $\beta$, typical walks in $\lambda_\text{max}$ are trapped in the vicinity of $\lambda=-1$. Thus all networks are SF. As $\beta$ is lowered, some walks escape the neighbourhood of $\lambda=-1$ and give rise to a second peak in the $\lambda_\text{max}$-distribution. As $\beta$ is further decreased more and more walks escape, thus shifting the balance towards the  less stable non-SF networks. Simultaneously, the smaller $\beta$ the farther away from $\lambda=-1$ typical walks get.  Finally, the choice of a very low $\beta$ allows all walks to escape from $\lambda=-1$ and get close to $\lambda=0$, resulting in a dominating peak of less stable networks, which are exponential.

{\it Finding key mechanisms--} It appears of interest to find which of the steps (i)-(iii) prove necessary to construct SF networks in the above way. For this we relax constraints in our construction procedure and study networks that result therefrom.

We have already noted, that a high value of $\beta$ is required to obtain SF networks. Since, via (\ref{Eq_acc}), $\beta$ determines the chance that a positive step in the $\lambda$-walks is accepted, a high $\beta$ could also be interpreted as a trap confining walks to $\lambda=-1$. Indeed, it turns out, that Eq. (\ref{Eq_acc}) can be replaced by a sharp cut-off as $p_\text{accept}=\Theta(\lambda_\text{max}(N)-\lambda_1)$, $\lambda_1 <\lambda_\text{c}$, still resulting in SF networks. One can conclude, that the functional form of Eq. (\ref{Eq_acc}) is not essential. To construct $S_-$ it is only required to keep the eigenvalue distribution `narrow' and `close' to $\lambda=-1$.

As a next key element in our construction procedure we included random network growth. To check the importance of this step, we investigate the influence of optimization for a narrow eigenvalue distribution alone. More specifically, we consider an uncorrelated random network of size $N$ with $L=4(N-1)$ links, of which half are negative and half positive. By swapping links between arbitrary nodes new network configurations are suggested, which are again rejected/accepted on the basis of step (iii) and Eq. (\ref{Eq_acc}). Even choosing very small network sizes ($N=20$) after $10^7$ iterations we still find peaked degree distributions as in the initial random network. Hence, our simulations suggest that this procedure does not lead to the formation of SF networks. Thus, as in preferential attachment, growth seems to be a key requirement to obtain SF networks.

{\it Drawing link strength at random--} So far we identified two necessary steps to produce SF graphs from the above matrix stability criterion: a tendency to trap eigenvalues close to $\lambda=-1$ and network growth. However, since we only allow links of strength $+1$ and $-1$ the procedure still lacks generality. In the following we draw link strengths from uniform distributions over the intervals $[-1,0]$ and $[0,1]$. Hence we now have negative links of strength $-1\leq s\leq 0$ and positive links in $0\leq s\leq 1$, allowing for a discrepancy $\langle m_{i j}\rangle_{i\not= j} \not= 0$ between the total weight of positive and negative links. In this way we construct an ensemble of {\it weighted} directed graphs 

As for the graphs with link strength $s=\pm 1$ we again find a bimodal distribution of $\lambda_\text{max}(N)$. Choosing a suitable $\lambda_\text{c}$ a more stable ($S_-$) and a less stable ensemble ($S_+$) of graphs can be defined as previously.
Again, networks belonging to the more stable ensemble are SF (with exponents $\gamma_\text{in}=\gamma_\text{out}=-2.35\pm 0.04$ for $N=100$ and $\beta=50$). Similarly, graphs in the less stable ensemble approach an exponential degree distribution for small $\beta$.

\begin{figure} [tbp]
\begin{center}
\includegraphics [width=7cm]{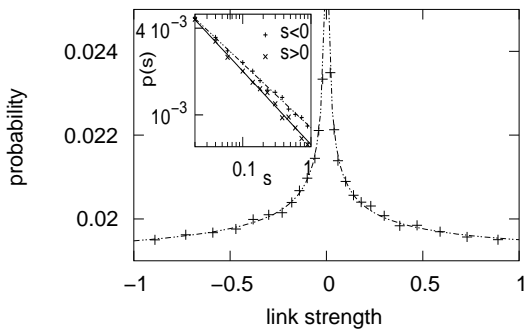}
 \caption {Distribution of the strengths of links for the more stable subset $S_-$ for networks of size $N=50$ constructed with $\beta=50$ (points). Interestingly, the more stable graphs have more strong links than the less stable ones (not shown). In the ensemble of more stable graphs the distribution follows two distinct power laws $Pr(s)\sim s^{-\delta_{-/+}}$ (dashed lines), with exponents $\delta_+=.51\pm 0.03$ ($s>0$) and $\delta_-=.45\pm 0.02$ ($s<0$). The inlet shows the different power laws holding for $s>0$ and $s<0$ on log-log scales. The data are logarithmically binned and represent averages over $10^5$ independent runs.}
\label{F6}
\end{center}
\end{figure}

To proceed, we examine the distribution of link strength of the more stable graphs $S_-$ [see Fig. \ref{F6}].
 Generally, as would be expected, weak links are much more frequent than strong links. Further, in the case of the ensemble of more stable matrices one finds distinct power laws  $Pr(s)\sim s^{-\delta_{-/+}}$ for both the negative and the positive branch of the distribution. For $s<0$ an exponent $\delta_+=.51\pm 0.03$ is obtained, while for $s<0$ it holds $\delta_-=.45\pm 0.02$. A similar behaviour, i.e. dominance of relatively weak links and the existence of only a few strong links (`hot-links'), is expected in many empirical networks in biology \cite{Bar-NatureGenetics}. Further, a recent empirical study confirms power law behaviour in the link strength distribution of some SF networks \cite{Barrat}.


Most of the empirically investigated networks are found to be not only SF, but to also show a high amount of `cliquishness' \cite{handbook}. This is measured by $c$, the clustering coefficient, which calculates the fraction of links between neighbours' neighbours which are actually present. In order to factor out the effects of a specific degree distribution, the values $\langle c\rangle$ found in the ensemble $S_-$ are compared to averages over an ensemble of randomized graphs $\langle c_\text{rand}\rangle$ with the same degree distribution (cf. \cite{Maslov}) and to random networks of the Erd\"os-R\'enyi type \cite{ER} ($c_\text{ER}$). For $\beta=50$, $N=100$ and link strength $|s|$ randomly drawn from $[0,1]$ our simulations result in $\langle c\rangle=.078$ to be compared with $\langle c_\text{rand}\rangle= .045$ and $c_\text{ER}=.025$. Hence, networks in the stable ensemble are substantially more cliquish than random networks. Preliminary studies seem to indicate only a slight dependence on $\beta$ and a ratio $\langle c\rangle/\langle c_\text{rand}\rangle$ which grows with the system size $N$. Similar experiments for the average shortest path length give values very close to that found in random networks, hinting to a small world like topology of our graphs.


In summary, using a stability criterion based on the largest real part of eigenvalues of a matrix associated with a graph, we have presented a model to generate an ensemble of graphs distinguished by their stability. In this context, more stable graphs turn out to have SF degree distributions. We identified two key mechanisms to obtain SF graphs in this way: growth and a tendendency to keep the real part of the eigenvalue distribution `narrow'. Extending the model to generate weighted graphs, we find stable graphs that are SF and exhibit a power law distribution of the link strengths. We find that the networks are substantially more cliquish than random networks, while still exhibiting a path length very similar to random networks. The model thus for the first time relates SF `small world' like graphs to a stability criterion which is directly associated with the underlying equations' dynamics and thus adds another notion to the understanding of SF networks as networks distinguished by robustness against perturbations.


\begin{thebibliography}{99}
\bibitem{handbook}{ {\it Handbook of Graphs and Networks}, edited by S. Bornholdt and H.~G. Schuster (Wiley-VCH, Berlin, 2002); R. Albert and A.-L. Barab\'{a}si, Rev. Mod. Phys. {\bf 74}, 47 (2002).}
\bibitem{Bar1}{A.-L. Barab\'{a}si and R. Albert, Science {\bf 286}, 509 (1999).}
\bibitem{Bianconi}{G. Bianconi, A.-L. Barab\'{a}si, and R. Albert, Europhys. Lett.  {\bf 54}, 436 (2001).}
\bibitem{Sole}{R. Ferrer and R.~V. Sol\'{e}, {\it Optimization in Complex Networks} in: {\it Statistical Physics of Complex Networks, Lecture Notes in Physics}  (Springer, Berlin, 2004).}
\bibitem{Berg}{J. Berg and M. L\"assig, Phys. Rev. Lett. {\bf 89}(22), 228701 (2002).}
\bibitem{Capocci}{A. Capocci, G.Caldarelli, R. Marchetti, and R. Pietronero, Phys. Rev. E. {\bf 64}, R035105 (2001).}
\bibitem{Sneppen}{B.J. Kim, A. Trusina, P. Minnhagen, and K. Sneppen, preprint: arXiv:nlin.AO/0403006 (2004).}
\bibitem{Bar2}{R. Albert, H. Jeong, and A.-L. Barab\'{a}si, Nature {\bf 406}, 387 (2000).}
\bibitem{May}{R.~M. May, Nature {\bf 238}, 413 (1972).}
\bibitem{OIKOS}{C.~C. Wilmers, S. Sinha, and M. Brede, OIKOS {\bf 99}, 363 (2002).}
\bibitem{Redner}{P.~L. Krapivsky and S. Redner, Phys. Rev. Lett. {\bf 85}, 4629 (2000).}
\bibitem{Maslov}{S. Maslov and K. Sneppen, Science {\bf 296}, 910-913 (2002).}
\bibitem{ER}{P. Erd\"os and A. R\'enyi, Publ. Math. Inst. Hung. Acad. Sci. {\bf 5}, 17-61 (1960).}
\bibitem{Barrat}{A. Barrat, M. Barth\'{e}l\'{e}my, R. Pastor-Satorras, and A. Vespignani, preprint: cond-mat/0311416 (2003).}
\bibitem{Bar-NatureGenetics}{A.-L. Barab\'asi and Z. N. Oltvai, Nature Genetics {\bf 5}, 101 (2004).}
\end{thebibliography}
\end{document}